\documentclass[sigconf]{acmart}

\AtBeginDocument{%
  \providecommand\BibTeX{{%
    \normalfont B\kern-0.5em{\scshape i\kern-0.25em b}\kern-0.8em\TeX}}}

\makeatletter
\renewcommand\@formatdoi[1]{}
\makeatother
\begin{document}

\title{Semantic In-Domain Product Identification for Search Queries}


\author{Sanat Sharma}
\affiliation{%
  \institution{Adobe Inc.}
  \city{San Jose}
  \state{California}
  \country{USA}}
\email{sanatsha@adobe.com}

\author{Jayant Kumar}
\affiliation{%
  \institution{Adobe Inc.}
  \city{San Jose}
  \state{California}
  \country{USA}}
\email{jaykumar@adobe.com}

\author{Twisha Naik}
\affiliation{%
  \institution{Adobe Inc.}
  \city{San Jose}
  \state{California}
  \country{USA}}
\email{tnaik@adobe.com}

\author{Zhaoyu Lu}
\affiliation{%
  \institution{Adobe Inc.}
  \city{San Jose}
  \state{California}
  \country{USA}}
\email{lolu@adobe.com}

\author{Arvind Srikantan}
\affiliation{%
  \institution{Adobe Inc.}
  \city{San Jose}
  \state{California}
  \country{USA}}
\email{asrikantan@adobe.com}

\author{Tracy Holloway King}
\affiliation{%
 \institution{Adobe Inc.}
 \city{San Jose}
 \state{California}
 \country{USA}}
 \email{tking@adobe.com}

\renewcommand{\shortauthors}{Sharma, et al.}

\begin{abstract}
 Accurate explicit and implicit product identification in search queries is  critical for enhancing user experiences, especially at a company like Adobe which has over 50 products and covers queries across hundreds of tools.  In this work, we present a novel approach to training a product classifier from user behavioral data.  Our semantic model led to: $>$25\% relative improvement in CTR (click through rate) across the deployed surfaces; a $>$50\%  decrease in null rate; a 2x increase in the app cards surfaced, which helps drive product visibility.
\end{abstract}



\maketitle

\section{Introduction}
\label{sec:intro}

\begin{figure}[htb]
    \centering
    \includegraphics[width=3.2in]{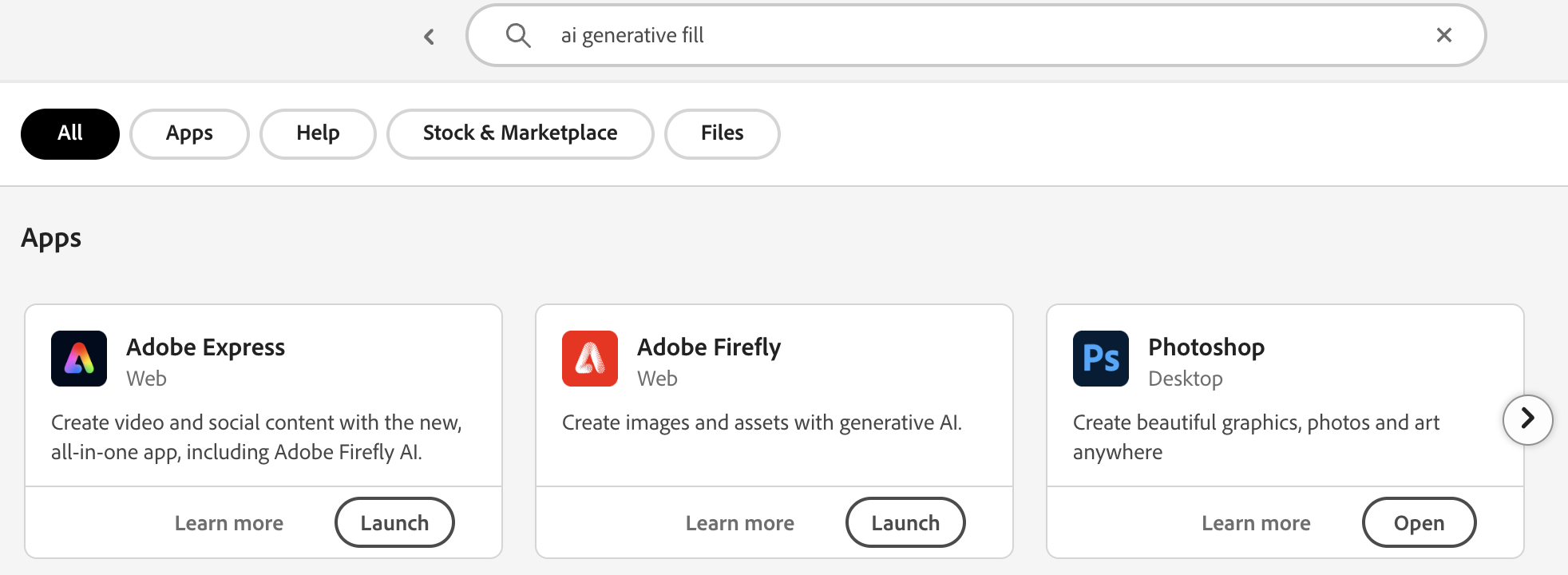}\\[2ex]
    \includegraphics[width=3.2in]{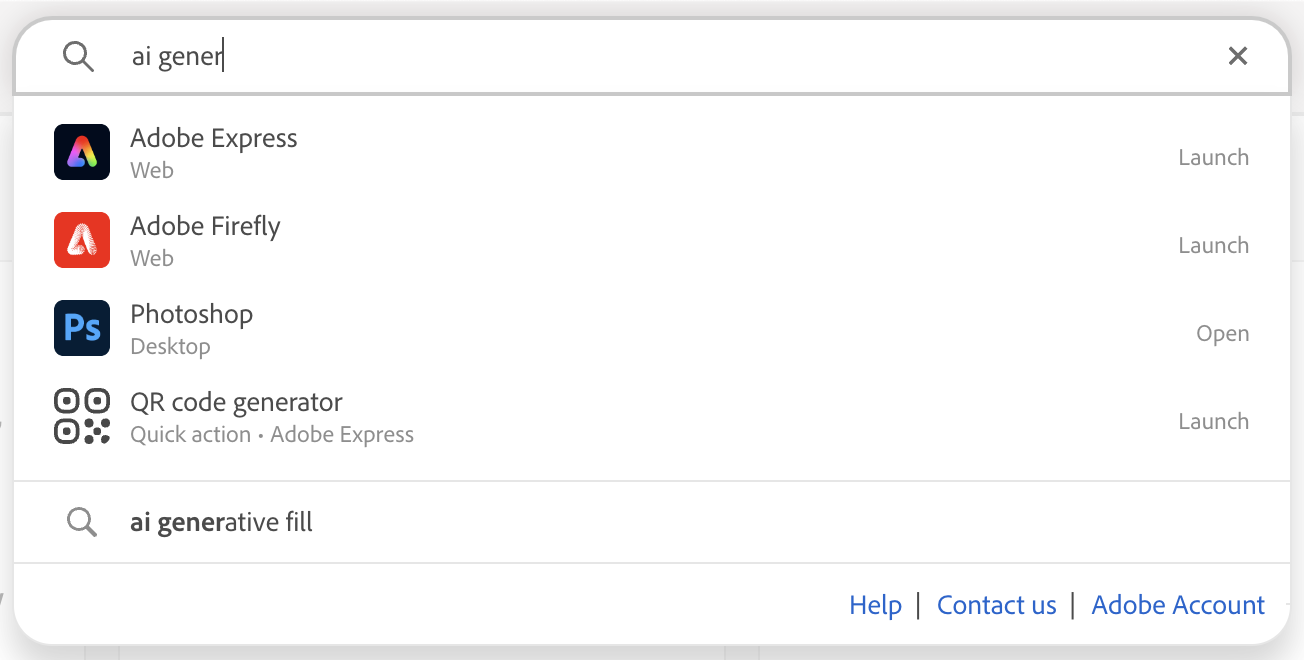}
    \caption{Product App Card Experiences: Top: App cards at the top of search results for \textit{ai generative fill}. Bottom: Autocomplete for  \textit{ai genera} with textual query suggestions are shown below the app cards. The product intent is implicit.}
    \label{fig:example}
    \end{figure}

Adobe boasts over 50 products for a variety of creative use cases (e.g.\ editing photos, videos, and audio, creating illustrations, animations, and vector graphics). When users come to Adobe.com or to Creative Cloud (CC, a subset of Adobe products focused on creativity), it is critical to route them to the right product for their use case. This is primarily done through contextual app card suggestions in autocomplete and at the top of search results. App cards provide users an easy way to discover, learn more about, or simply launch the Adobe product that matches their query intent. These app cards are the most clicked items on app-agnostic surfaces like CC and Adobe.com and are critical in driving new-user acquisition and product discovery, as well as providing existing users with help for their queries. Example app card triggering in autocomplete for implicit product intent is shown in Figure \ref{fig:example}.

Initially, this matching was done via regular expression rules and simple named entity recognition. As a result, minor variations in phrasing would result in app cards not triggering and queries with implicit product intent (e.g.\ \textit{edit video} should trigger app cards for Premiere Pro and Rush) rarely triggered app cards. In this work, we describe a  low-latency query-to-product semantic matching system that now provides contextual app card suggestions for the search and autocomplete services.

\section{Prior Art}

Product disambiguation has often been modeled as a (named) entity extraction (NER) approach in industry. Most approaches to query product disambiguation have been to model tokens within queries and extract relevant entities based on the set of supported products \cite{Luiggi_2023,yamada-etal-2022-global}. Recent works have also leveraged autoregressively re-writing the query for easier disambiguation and then using a retrieval or classification head on top \cite{DBLP:journals/corr/abs-2010-00904}. There has also been recent research for NER tagging in low-resource cases, where named entities are specialized \cite{DBLP:journals/corr/abs-2112-00405}, which is the case for Adobe product intent. 

In addition to NER techniques, other works have focused on a semantic search approach of mapping product embeddings and query embeddings in the same semantic space \cite{DBLP:journals/corr/abs-1907-00937}. 

Our work utilizes a mixture of components from previous approaches. We pretrain a language model (LM) on our internal document set to learn the  intricacies of Adobe products \citep{kumaretal-sirip23} and then utilize a classification approach on top of the LM for product matching. We found this approach to work better than semantic search for products which are less frequently referenced in queries and less popular with users due to their highly specific applications (e.g.\ Adobe Bridge) and much better than NER in cases where no products are explicitly mentioned in the user query (e.g.\ \textit{redact document} for Acrobat and \textit{edit video} for Premiere Pro and Rush).

\section{Datasets}
\label{sec:dataset}

We support 46 Adobe products  in our training dataset. The datasets are in English but come from multiple locales since English queries are used in combination with non-English in most locales. In order to learn a good representation of the products and  to tackle a diverse set of queries, we utilize four datasets, from user clicks to expert-maintained spreadsheets. These are described below.

\textbf{Adobe HelpX Behavioral Dataset}
Adobe HelpX\footnote{\url{helpx.adobe.com}} gets millions of unique visitors every year looking for tutorials and learn content (Adobe HelpX articles, Adobe help videos) related to Adobe products. We utilize the click logs from user queries $\rightarrow$ HelpX article clicked to generate our query $\rightarrow$ product dataset. From each help article, we extract the related product from the metadata. We use logs from January 2021 -- August 2022 for our training set. This dataset is noisy but provides a large and diverse dataset that is critical for learning a good representation.
A sample row from the dataset is shown in Table \ref{table:trainingrow}.

\begin{table}[htb]
\caption{Example training data with the query, clicked document, associated product, and log click ratio score}
\begin{tabular}{|l|l|}
\hline
\small 
\textbf{Field} & \textbf{Example Value}\\\hline\hline
{Query} & change color of text\\\hline
{Document} & \url{https://HelpX.adobe.com/indesign/}\\
& \url{using/editing-text.html} \\\hline
{Product} & Acrobat\\\hline
{Relevance} & 0.24\\
{Log click ratio} &\\
\hline
\end{tabular}

\label{table:trainingrow}
\end{table}

We  utilize a relevance field that is derived by using the log of the click ratio of the query-document pair. This is important because for a given query, there may be multiple clicked documents  and we wish to pay more attention to higher click pairs. Since we can show multiple app cards to the user (Figure \ref{fig:example}, the applications use the relevance scores both to determine which app cards to shown and their order. We take the log of the max click ratio to allow less frequently clicked documents to  be part of the learning process.

\begin{equation}
    relevance = log(\frac{clicks(q_i \longrightarrow d_j)}{max(clicks(q_i \longrightarrow D_i))})
    \label{eq:rel}
\end{equation}

    \noindent where $q_i$ represents the specific query; 
    $d_j$ represents the particular document clicked;
  $clicks$ represents the number of clicks for the pair; 
     $D_i$ represents the set of documents clicked for $q_i$, ie $D_i = \{d_i...d_n\}$.

\textbf{HelpX Document Dataset}
Adobe HelpX documents are curated by Adobe content creators to provide information about Adobe products, from tutorials to product announcements. We  utilize this high-quality resource by considering the document title and description as unique query-product training pairs. Since this dataset is curated and high quality, it is given a higher weight during training (relevance = 1).

\textbf{Product NER Explicit Dataset}
From user queries in Creative Cloud, we utilize a rule-based product NER to extract query-product pairs containing explicit product names. This dataset allows us to train on a wide set of explicit, high precision intent queries.

\textbf{Adobe Express Dataset}
Adobe Express is one of the newer Adobe products  and hence has few user clicks in our behavioral datasets. To bolster additional training data for this product and to learn a good representation for its queries, we utilize top Express in-product queries as part of our dataset.

Finally, we merged  the four datasets. Each unique query may have 1 or more products assigned to it. The overall dataset counts are shown in Table \ref{tab:datasize}.

\begin{table}[htb]
\caption{Dataset Size: Size is in number of rows of query-document-product-click ratio (see Table \ref{table:trainingrow})}
\begin{tabular}
{|l|l|}
\hline
\small 
\textbf{Dataset} & \textbf{Unique Rows} \\
\hline
Adobe HelpX Behavioral Dataset & 177500 \\
Adobe HelpX Document Dataset & 11757 \\
Adobe Express Dataset & 6637 \\
Product NER Explicit Dataset & 5208 \\
\hline
\end{tabular}

\label{tab:datasize}
\end{table}

\section{Model}

We divide the task of learning a good representation of user queries into two parts. The first part is pretraining a language backbone to utilize for downstream finetuning. The second is training the classifier head on top of the language backbone.

\subsection{Language Model Pretraining}

\begin{figure*}[htb]
    \centering
    \includegraphics[width=5in]{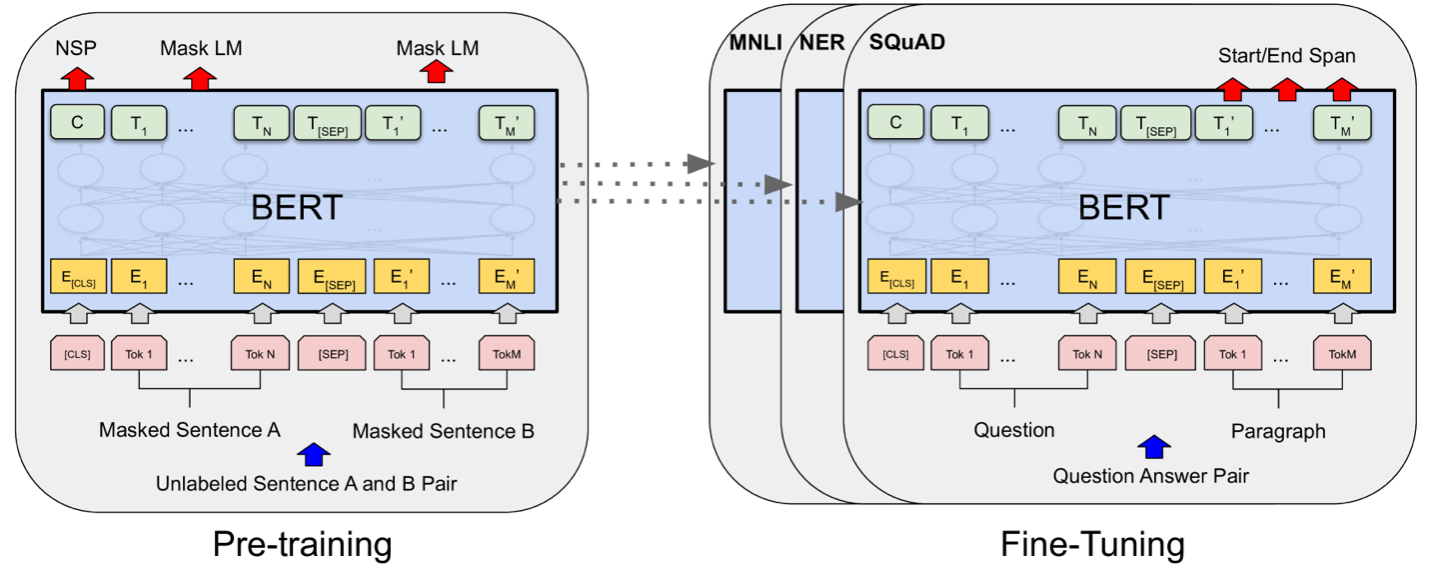}
    \caption{DeBERTa Pretraining: We break  HelpX documents into blocks of 128 tokens and pretrain. This allows the LM to understand Adobe product vocabulary and features better.}
    \label{fig:model}
\end{figure*}

We found open-source language models (LMs) like BERT \cite{DBLP:journals/corr/abs-1810-04805} to be inadequate for Adobe user queries (see also \cite{kumaretal-sirip23} on training an Adobe-specific language model for semantic search). This is primarily due to two reasons:
\begin{enumerate}
    \item \textbf{Lack of knowledge of Adobe products}: open-source models are trained on general web data and do not understand the intricacies of Adobe products. Product features such as crop or generative fill were not understood by the general models. In addition, some product names (e.g.\ Illustrator, Rush) are also common English words).
    \item \textbf{Poor product disambiguation}: We found open-source models to have trouble disambiguating products with similar names. Products like Premiere Pro and Premiere Rush or Photoshop and Photoshop Express were lumped together despite being unique products.
\end{enumerate}

To counter this, we pretrained a LM based on Microsoft DeBERTa v3 \cite{DBLP:journals/corr/abs-2111-09543} starting from publicly available pretrained weights on the HelpX document dataset using masked language modeling techniques. We arbitrarily split our Adobe-specific datasets (section \ref{sec:dataset}) into training and validation sets. We trained on block sizes of 128 and found the model to showcase good perplexity. Perplexity  in language modeling gauges how well a probability model predicts a sample. See Table \ref{tab:trainingdetails} for a summary.

\begin{table}[htb]
\caption{Training Details}
\begin{tabular}{l}
\hline
Break HelpX documents into blocks (128 tokens)\\
Concatenate\\\hline
Train size: 107240 examples\\
Validation size: 5645\\
Trained model perplexity: \textbf{7.47}\\
\hline
\end{tabular}
\label{tab:trainingdetails}
\end{table}

Pretraining the LM on HelpX data causes a \textbf{14\%} improvement in downstream classification accuracy compared to using a pretrained LM. This reinforced our hypothesis that domain-specific workflows such as Adobe help content have a different data distribution than open datasets.

\subsection{Classifier Training}

Once we trained our LM backbone, we trained a classifier to predict Adobe products given a user query. We utilized the  training datasets described in section \ref{sec:dataset} for this classifier head. We experimented with freezing the LM backbone (no weights are updated in the LM) and found the best combination to be to freeze the backbone for the initial few epochs and then train the full system for a few additional epochs.

We utilized a classic 2 hidden-layer Multilayer Perceptron network, with 0.5 dropout rate and a learning rate of 1e$-$5 and trained the classifier in a multi-label approach, i.e.\ each product was given a probability score between 0--1 given a query. The multi-label approach is necessary  because a large-number of implicit product queries are associated with multiple products and even explicit product queries  can be associated with multiple products (e.g.\ \textit{photoshop} is primarily associated with three Photoshop products (web, mobile, and desktop) as well as with Photoshop Express).
We use Weighted Binary Cross Entropy loss function for our training and leverage the relevancy weights (see equation \ref{eq:rel}) to pay more attention to more important examples during training.

\section{Offline Evaluation and AB Testing}

\subsection{Quantitative Evaluation on Behavioral Queries}
We reserve 10\% of our initial dataset (section \ref{sec:dataset}) for  evaluation. We compute per-product and per-source metrics. Since the dataset  comprises past user queries,  it reflects the final product use cases (see section \ref{sec:intro} and figure \ref{fig:example}). However, it is focused on explicit product mentions since the production app cards are primarily triggered for explicit mentions. Even with explicit product intent, given query may have multiple products associated with it based on past user click behavior; each of these is considered in the quantitative evaluation. As shown in Figure \ref{tab:evalquant}, precision and recall are well balanced and result in an F1 score of .949.

\begin{table}[htb]
\caption{Quantitative Evaluation}
\begin{tabular}{|c|c|c|c|c|}
\hline
\multicolumn{5}{|c|}{Quantitative Metrics on the Testset} \\\hline
\textbf{Rows} & \textbf{Precision} & \textbf{Recall} & \textbf{Accuracy} & \textbf{F1} \\
22849 & .961 & .941 & .970 & .949\\
\hline

\end{tabular}
\label{tab:evalquant}
\end{table}

\subsection{Qualitative Manual Annotation of Implicit Intent}

Through quantitative evaluation on the test set (previous subsection), we determined that the model did well when the product was mentioned in the query. We then focused on queries with implicit information about the product, e.g.\ {\em keyframe caddy}, {\em fashion poster}, etc. We utilized a set of 2700 production CC queries  for evaluation. These queries were previously unseen by the model and we utilized Adobe-internal product experts (e.g.\ product managers for CC) to  judge relevancy of the predicted product to the user query.
For each query, the model predicts the most likely output(s). Then the product experts  mark the suggested output as correct/relevant or incorrect/irrelevant.
In the cases where multiple products were predicted, the evaluators were asked to mark the predications as correct/relevant only when all the products predicted were useful. That is, the entire product intent prediction from the model had to be correct, not just a subset of the predictions. Table \ref{tab:evalqual} shows the accuracy results for the 2700 queries in the qualitative evaluation.

\begin{table}[htb]
\caption{Qualitative Metrics  for Implicit Product Queries: The entire set of intents for a given query must be correct to count as correct.}
\begin{tabular}{|c|c|c|c|}
\hline
\multicolumn{4}{|c|}{Qualitative Metrics for Implicit Product Queries}\\\hline
\textbf{Rows} & \textbf{Correct} & \textbf{Incorrect} & \textbf{Accuracy} \\
2700 & 2452 & 181 & \textbf{.931} \\
\hline
\end{tabular}
\label{tab:evalqual}
\end{table}

\subsection{AB Testing}
We AB-tested the new product intent model for showing app cards in autocomplete and at the top of search results. The new model was tested for all locales against the production model on the CC app and CC web site. Although the LM is trained for English queries, the large number of English queries in non-English locales and the fact that Adobe product names are identical in all languages means that the model triggers app cards in all locales.

As hypothesized, app cards surfaced significantly more due to triggering on implicit product intent queries. Previously, a few fixed queries and key terms triggered app cards, but there was no semantic model to provide broader coverage. The AB test showed a 2-fold increase in surfacing and  a $>$50\% decrease in queries with no app cards. We did not expect app cards for all queries since some queries do not have app intent; so, there should always be some queries without app cards surfacing.

Both autocomplete and search result app cards saw an increase in click-through rate for all surfaces and an increase in the unique users who engaged with app cards. Overall, there was a $>$25\% relative improvement in CTR (click-through rate) across the deployed surfaces.\footnote{We cannot share exact CTRs and so only include relative improvement.}\ These increases reflect the fact that the increased surfacing, e.g.\ for queries with implicit product intent, was high precision and provided information users needed.

\section{Conclusion and Future Work}

Accurate product identification is  critical for enhancing user experiences, especially at a company like Adobe which has over 50 products.  In this work, we present a novel approach to training a product classifier from user behavioral data.  Our semantic model led to: 

\begin{itemize}
    \item $>$25\% relative improvement in CTR (click-through rate) 
    \item a $>$50\%  decrease in null rate
    \item a 2x increase in the app cards surfaced, which helps drive product visibility.
\end{itemize}

As future work, to better support non-English queries with implicit product intent (e.g.\ {\em images gratuites} (French: `free pictures') which is associated with the Adobe Stock image marketplace), we are training a multi-lingual version of the model. In addition, we are also experimenting with better long prompt understanding for product disambiguation. This is particularly important for RAG based systems when dealing with retrieval for long prompts\cite{sharma2024retrieval}.

\bibliographystyle{ACM-Reference-Format}
\bibliography{sirip}

\end{document}